\documentclass[prb,twocolumn,showpacs,floatfix,aps]{revtex4}
\usepackage{graphicx}

\begin{document}

\title{Phonon structure in point-contact spectra of
MgB$_{2}$}
\author{I.K. Yanson}

\affiliation {B.I. Verkin Institute for Low Temperature Physicsa
and Engineering, National Academy of Sciences, 47 Lenin Ave.,
61103, Kharkiv, Ukraine.}

\email{yanson@ilt.kharkov.ua}

\date{\today}
\begin{abstract}
In strong-coupling superconductors with a short electron mean free
path the self-energy effects in the superconducting order
parameter play a major role in the phonon manifestation of the
point-contact spectra at above-gap energies. We compare the
expressions for the nonlinear conductivity of tunnel, ballistic,
and diffusive point-contacts and show that these expression are
similar and correspond to the measurements of the phonon structure
in the point-contact spectra for the $\pi $-band \ of MgB$_{2}$.
\end{abstract}

\maketitle

\section{Introduction}

The inelastic point-contact spectra of phonons in metals are based
on expansion of the nonlinear $I-V$ characteristic in terms
proportional to $\ d/l_{\varepsilon }$, where $d$ is the
characteristic size of metallic nanoconstriction connecting two
bulk metal half-spaces and $l_{\varepsilon }$ is the inelastic
electron mean free path \cite{KOS,Yan2000}.

If $d$ is smaller than the electron mean free path $l$ \ ($%
l^{-1}=l_{i}^{-1}+l_{\varepsilon }^{-1}$, $l_{i}$ is the elastic
mean free path), the electrons flow through the contact
ballistically. In opposite cases $(d\gg l)$ they perform a
diffusive motion, but if the energy relaxation length $\Lambda
_{\varepsilon }\approx \sqrt{l_{i}l\varepsilon }$ remains larger
than $d$, their movement inside the constriction is conservative,
and the inelastic point-contact spectroscopy is still possible
\cite{KulYan,KSS}.

In the superconducting state a new characteristic length appears:
the size of the Cooper pair or the coherence length $\xi ,$ and
the additional nonlinear \ features emerge being proportional to
the expansion of the so called ''excess'' current in terms
proportional to $d/\xi .$ In all the cases considered, the size of
the contact is smaller than the corresponding characteristic
length scale: $\xi ,\,l_{\varepsilon },\Lambda _{\varepsilon } $
\textit{etc}.

The zero-approximation in expansion of $I(V)$ in $d/l_{\varepsilon
}$ is the ohmic current considered by Sharvin \cite{Sha}. From the
Sharvin's formula the characteristic size $d$ of the contact can
be determined in the ballistic limit. The second derivative of the
first approximation in expansion of $I(V)$ in $d/l_{\varepsilon }$
is directly proportional to the spectral function of
electron-phonon interaction (PC EPI) $\ g_{PC}(\omega )=\alpha
_{PC}^{2}\left( \omega \right) F\left( \omega \right)$ of the
specific point-contact transport both in the normal and in the
superconducting states \cite{KOS,OmeKhl,Khl}. This term is the
basis of the canonical inelastic point-contact spectroscopy (PCS).
Here, $\alpha _{PC}^{2}\left( \omega \right) $ is the average
electron-phonon matrix element taking into account the kinematic
restriction imposed by contact geometry and $F\left( \omega
\right) $ is the phonon density of states.

In diffusive point contacts $(d\gg l_{i})$ the role of the scale,
where the
backscattering inelastic processes become essential, turns from $d$ in $%
l_{i} $ and in the case when $l_{i}/l_{\varepsilon }\ll 1$ the
corresponding contribution becomes negligible \cite{KulYan,KSS}.
This property will be essential when we consider the phonon
structure in point-contact spectra of dirty MgB$_{2}$ contacts in
the \textit{c}-direction.

In the superconducting state, due to the Andreev reflections of
quasiparticles at the $N-S$ boundary, the excess current appears
both in $S-c-N$ and $S-c-S$ contacts ($c$ stands for
constriction). Its magnitude is the constant proportional to
$\Delta $ for the weak-coupling BCS approximation at $eV\gg \Delta
$ ($\Delta $ is the superconducting energy gap)
\cite{ArtVolZai,BTK}. In the strong-coupling model of EPI the
energy gap depends on the energy of quasiparticles due to phonon
retardation effects reflecting the characteristic features of the
phonon spectrum at singular van-Hove energies. Hence, even in the
zeroth order the nonlinearities appear in $I(V)$ \cite{OmeBelKul}.
For the issue of this study it is most important that these
nonlinearities are independent of the elastic mean free path
(i.e., purity of the contact). In this connection one should keep
in mind that for the simple method of a gentle touch between two
sharp edges of the electrodes it is not possible to control the
elastic mean free path in the constriction. In what follows we
show that in spite of this restriction, the satisfactorily
reproducible results are obtained not only for the shapes of
phonon structure but also for their amplitude.

To conclude this section, we mention that  the use of the first
approximation in expansion of excess current in powers of $d/\xi $
allows the nonequilibrium effects in the vicinity of the
constriction to be developed, which are out of the scope in this
paper \cite{BelOme}.

\section{Theoretical background of zeroth-order elastic component in excess
current}

In this section we review what has recently become known about the
phonon structure in the point-contact spectra \cite{Yan2004}. In
order to do this, we should refer to the works where the excess
current in point contact is considered in terms of the
strong-coupling theory. That was first done in the paper of \
Omel'yanchuk, Beloborod'ko and Kulik for ballistic $S-c-N$ point
contact \cite{OmeBelKul}. \ The first derivative of
$I(V)$-characteristic at $T=0$ has the following form in this
case:

\begin{equation}
R_{N}\left( \frac{dI}{dV}\right) _{bal}=1+\left| \frac{\Delta
\left( \varepsilon \right) }{\varepsilon +\sqrt{\varepsilon
^{2}-\Delta ^{2}\left( \varepsilon \right) }}\right|
^{2},\hspace{0.2cm}\varepsilon =eV  \label{bal1}
\end{equation}
where $R_{N}$ is the resistance of the contact in the normal
state. This expression implies $\ d\ll \xi $ to be fulfilled ($\xi
$ is the coherence length in pure metal).

For diffusive point contact with the restriction: $l_{i}\ll d\ll \sqrt{%
l_{i}\xi },$ the same quantity takes the form \cite{Yan2004}:

\begin{equation}
R_{N}\left( \frac{dI}{dV}\right) _{dif}=\frac{1}{2}\ln \left| \frac{%
\varepsilon +\Delta \left( \varepsilon \right) }{\varepsilon
-\Delta \left(
\varepsilon \right) }\right| \frac{\Re\frac{\varepsilon }{\sqrt{%
\varepsilon ^{2}-\Delta ^{2}\left( \varepsilon \right) }}}{\Re\frac{%
\Delta (\varepsilon )}{\sqrt{\varepsilon ^{2}-\Delta ^{2}\left(
\varepsilon \right) }}},\hspace{0.2cm}\varepsilon =eV \label{dif1}
\end{equation}

We would like to compare these dependences with the expression for
the conductance of the tunnel junction due to the self-energy
effect in the superconducting energy gap:

\begin{equation}
R_{N}\left( \frac{dI}{dV}\right) _{tun}=\Re\frac{\varepsilon }{\sqrt{%
\varepsilon ^{2}-\Delta ^{2}\left( \varepsilon \right)
}},\hspace{0.2cm}\varepsilon =eV  \label{tun1}
\end{equation}

In the above mentioned formulae $\Delta \left( \varepsilon \right)
$ is the complex superconducting order parameter
\begin{equation}
\Delta \left( \varepsilon \right) =\Delta _{1}\left( \varepsilon
\right) +i\Delta _{2}\left( \varepsilon \right) ,  \label{gaps}
\end{equation}
where $\Delta _{1}\left( \varepsilon \right) $ and $\Delta
_{2}\left(
\varepsilon \right) $ are the $\Re \left( \varepsilon \right) $ and $\Im%
\Delta \left( \varepsilon \right) $, respectively. It would be
very instructive to express Eqs. (\ref{bal1}), (\ref{dif1}),
(\ref{tun1}) as functions of real and imaginary components.

Thus, using the auxiliary expression
\begin{equation}
 A\left( \varepsilon \right) =\sqrt{%
\left( \varepsilon ^{2}-\Delta _{1}^{2}+\Delta _{2}^{2}\right)
^{2}+4\Delta _{1}\Delta _{2}},
\end{equation}
the results are as follows:

\begin{eqnarray}
R_{N}\left( \frac{dI}{dV}\right) _{bal}=1+\frac{1}{\left( \Delta
_{1}^{2}+\Delta _{2}^{2}\right) } \nonumber\\ \times
\left[ \varepsilon ^{2}+A-\sqrt{%
2\varepsilon }\sqrt{A+\varepsilon ^{2}-\Delta _{1}^{2}+\Delta _{2}^{2}}%
\right] ,  \label{bal2}
\end{eqnarray}

\begin{eqnarray}
R_{N}\left( \frac{dI}{dV}\right) _{dif} =\frac{1}{2}\ln \left[ \frac{\sqrt{%
\left( \varepsilon ^{2}-\Delta _{1}^{2}-\Delta _{2}^{2}\right)
^{2}+4\varepsilon ^{2}\Delta _{2}^{2}}}{\left( \varepsilon -\Delta
_{1}\right) ^{2}+\Delta _{2}}\right]  \nonumber  \\
\times \frac{\varepsilon \sqrt{A+\varepsilon ^{2}-\Delta
_{1}^{2}+\Delta _{2}^{2}}}{\Delta _{1}\sqrt{A+\varepsilon
^{2}-\Delta _{1}^{2}+\Delta _{2}^{2}}\pm \Delta
_{2}\sqrt{A-\varepsilon ^{2}+\Delta _{1}^{2}-\Delta
_{2}^{2}}},\label{dif2}
\end{eqnarray}

where ''plus'' and ''minus'' in the last line stand for \ $\Delta
_{1}\Delta _{2}<0$ and  $\Delta _{1}\Delta _{2}>0$, respectively.

Finally, for tunnel junction the same quantity reads:

\begin{equation}
R_{N}\left( \frac{dI}{dV}\right) _{tun}=\frac{\varepsilon \sqrt{%
A+\varepsilon ^{2}-\Delta _{1}^{2}+\Delta _{2}^{2}}}{\sqrt{2A}}.
\label{tun2}
\end{equation}

Expressions (\ref{bal1}), (\ref{dif1}), (\ref{tun1}) \ [and (\ref{bal2}), (%
\ref{dif2}), (\ref{tun2})] seem quite different but in reality
they are very similar. This becomes evident if we compare their
asymptotic expressions for the limit $\varepsilon \gg \Delta $.
For three extreme regimes of current flow in aforementioned point
contacts the corresponding equations take the form:

for a ballistic point contact:

\begin{equation}
R_{N}\left( \frac{dI}{dV}\right) _{bal}\approx 1+\frac{\Delta _{1}^{2}}{%
4\varepsilon ^{2}}+\frac{\Delta _{2}^{2}}{4\varepsilon ^{2}},
\label{bal3}
\end{equation}

for a diffusive point contact:

\begin{equation}
R_{N}\left( \frac{dI}{dV}\right) _{dif}\approx 1+\frac{\Delta _{1}^{2}}{%
3\varepsilon ^{2}},  \label{dif3}
\end{equation}

and for a tunnel junction:

\begin{equation}
R_{N}\left( \frac{dI}{dV}\right) _{tun}\approx 1+\frac{\Delta _{1}^{2}}{%
2\varepsilon ^{2}}-\frac{\Delta _{2}^{2}}{2\varepsilon ^{2}}.
\label{tun3}
\end{equation}
In all these equations $\varepsilon =eV$ and $T=0$ are implied.

Since $\Delta _{2}\left( \varepsilon \right) $ becomes comparable
to $\Delta _{1}\left( \varepsilon \right) $ only at the highest
phonon energy, almost in the whole phonon energy range these three
extreme regimes of current flow
differ only little, mostly in amplitudes, which decrease in the tunnel$%
\rightarrow $diffusive$\rightarrow $ballistic contact row like
their prefactors $\frac{1}{2}\rightarrow \frac{1}{3}\rightarrow
\frac{1}{4}$. This derivation is of a key importance to
understanding the quite similar phonon point-contact spectra for
the different contacts randomly accessible in the experiment.

Let us illustrate the above theoretical formulae by an example of
superconducting Pb frequently used for these purposes in tunneling
spectroscopy of the electron-phonon interaction. This metal has
the relatively strong EPI and a simple phonon spectral function,
consisting of two acoustic peaks: transverse (TA) and longitudinal
(LA) (Fig. 1). We take the tabulated functions: $\alpha
^{2}(\varepsilon )F(\varepsilon )$, $\Re\Delta \left( \varepsilon
\right) $ and $\Im\Delta (\varepsilon )$ from
Rowell-McMillan-Dynes$\prime$ preprint \cite{RowDyn} and plot both
the EPI spectral function and the
second derivative characteristic $-\frac{d}{dV}\left( R_{N}\frac{dI}{dV}%
\right) $ proportional to the experimentally recorded second
harmonic of modulation voltage, $V_{2}(eV)$ (see the text below).
One can immediately see that three point-contact spectra (tunnel,
diffusive and ballistic) are very similar differing only in the
constant prefactor at $\left[ \frac{\Re\Delta \left( \varepsilon
\right) }{\varepsilon }\right] ^{2}$ (see above). The differences
in shape occur only at the second peak of the EPI spectrum, where
$\Im\Delta (\varepsilon )$ becomes essential. For the transverse
peak the positions of maxima of point-contact spectra $V_{2}(eV)$
coincide with the TA-maximum of EPI function shifted by $\Delta $
to the right ( we remind that throughout this paper $S-N$ contact
is considered). On the other hand, the maxima of $V_{2}(eV)$
functions at LA peak are slightly displaced with respect to the
EPI function and their shapes change, which becomes more pronouced
the further we deviate from the tunnel junction in the row
mentioned above.

\begin{figure}
\includegraphics[width=8cm,angle=0]{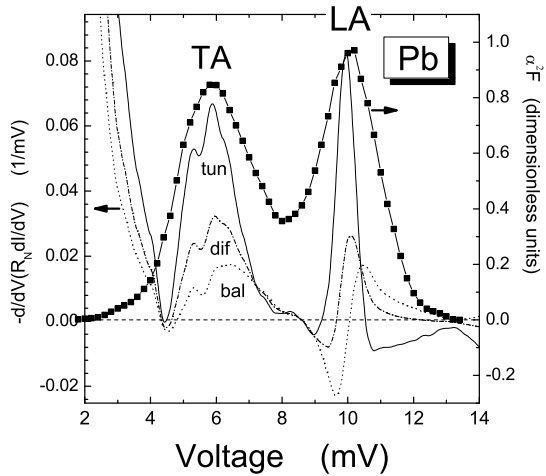}
\caption[]{The second derivative of $I-V$ characteristics for
superconducting
lead $S-c-N$ point contact calculated by Eqs. (\ref{tun2}), (%
\ref{dif2}), (\ref{bal2}) for the tunnel (solid curve), diffusive
(dash-dotted curve) and ballistic (dotted curve) regimes of
current flow. $\Re\Delta \left( \varepsilon \right) $\ and
$\Im\Delta \left( \varepsilon \right) $\ are taken from the
Rowell-McMillan-Dynes \cite{RowDyn} tabulation. Squared symbols
stand for Rowell-McMillan-Dynes \cite {RowDyn} EPI spectral
function shifted by $\Delta =1.4$ meV to the right on the voltage
scale. Its ordinate scale is shown on the right side.}
\label{fig1}
\end{figure}

Comparing the exact spectra (Eqs. (\ref{bal2}), (\ref{dif2}),
(\ref{tun2})) with the asymptotic ones (Eqs. (\ref{bal3}),
(\ref{dif3}), (\ref{tun3})), we see that for Pb the differences
are only minor (the example for diffusive point-contact is shown
in Fig. 2). It goes without saying that the differences for
materials with greater ratios of phonon energy to the energy gap,
$\hbar \omega _{ph}/\Delta $, will be very small. For example, in
Sn or MgB$_{2}$ , where $\hbar \omega _{ph}/\Delta \gg 1$ the
differences between exact and asymptotic curves in the phonon
energy range become indistinguishable.

\begin{figure}
\includegraphics[width=8cm,angle=0]{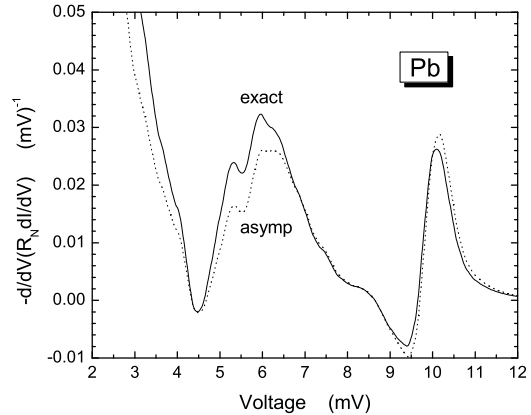}
\caption[]{Comparison of the exact (\ref{dif2}) and asymptotic
(\ref{dif3}) calculated curves for lead in the $S-c-N$ point
contact. } \label{fig2}
\end{figure}

Experimentally, the expected shift of the Pb phonon spectra to the
right was first noticed in Ref. \onlinecite{Khot}. Unfortunately,
in this paper (and also in the review paper \cite{Yan2000}) the
experimental curves were compared with theoretical ones without
allowing for the shift of the latter by $\Delta $ to the right. In
Fig. 3 this drawback is corrected and the important consequences
are drawn for the ballistic point contact. The upper panel of Fig.
3 displays two experimentally recorded dependences: i) the second
derivative of the $I\_V$ characteristic in the normal state
(superconductivity is destroyed by the magnetic field) and ii) the
same dependence in the superconducting state. The BCS dependence
in the superconducting state is shown for comparison. The
theoretically expected point-contact spectrum for ballistic
contact is shown in the bottom panel of Fig. 3. It is compared
with the experimental PC EPI spectrum in the normal state shifted
by $\Delta _{Pb}=1.4$ meV . As expected, the position of the
\textit{calculated} TA-maximum coincides with the TA-maximum
in the normal-state spectrum and the corresponding feature (zero of $y$%
-coordinate) coincides with the normal-state LA-maximum (see Fig.
1).

\begin{figure}
\includegraphics[width=8cm,angle=0]{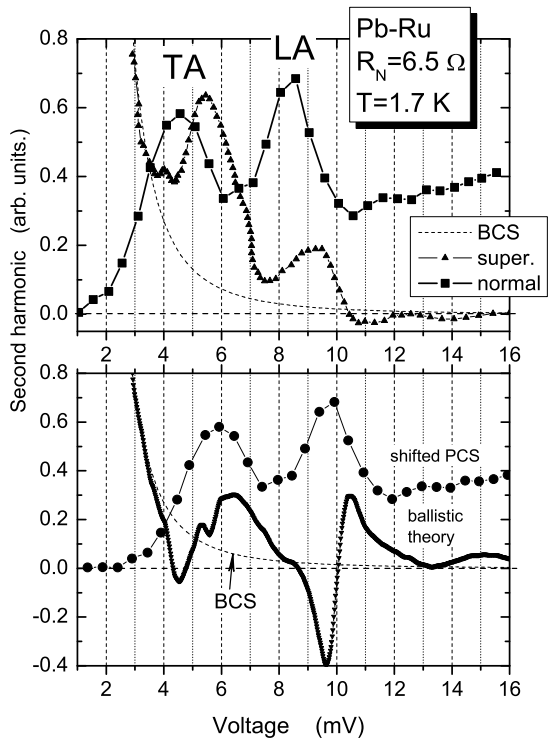}
\caption[]{The Pb-Ru experimental point-contact spectra in the
normal (squared symbols) and superconducting (triangles-up) states
(upper panel) \cite{Khot}. The calculated spectrum for a ballistic
point contact is shown in the lower panel (triangles-down)
together with the experimental PC EPI spectrum shifted by $\Delta
=1.4$ meV to the right. The theoretical BCS dependence is shown
for comparison (thin dotted curve).} \label{fig3}
\end{figure}

In contrast, the phonon features on the \textit{experimental}
curve in the superconducting state shown in the upper panel are
noticeably displaced to the left from the expected calculated
positions (shown in the lower panel) towards the non-shifted
phonon maxima. This displacement is the result of mixing at least
of two different mechanisms of revealing the phonon features:
namely, the inelastic backscattering current, which has the same
positions both in the superconducting and the normal states (the
latter is shown as an experimental PC\ EPI curve in upper panel of
Fig. 3), and the theoretical zeroth-order approximation of excess
current shown as a theoretical curve in the lower panel. Such
mixing seriously impedes interpretation of PC spectra of a
ballistic strong-coupling superconductor. Fortunately, as
mentioned above, in diffusive point contact the inelastic
backscattering current is negligible and the pure self-energy
effect on excess current becomes clearly visible. The latter
situation holds in MgB$_{2}$ films measured along the
$c$-direction as well.

\section{Self-energy effect in $c$-axis oriented MgB$_{2}$ films}

For two-band superconductor MgB$_{2}$ (see the review
\cite{MazAnt}) the $\Re\Delta \left( \omega \right) $ and
$\Im\Delta \left( \omega \right) $ for $\pi $ and $\sigma $-band
are depicted in the upper and lower panel of Fig. 4
\cite{Yan2004}, respectively. We are interested mostly in the $\pi
$-band dependences, since for thin films oriented along the
$c$-axis only the $\pi -$band part of the Fermi surface has
nonzero Fermi velocity for charge carriers (holes) along the
contact axis, playing the main role in the contact current. For
the $\sigma $-band, which is two-dimensional, only holes with the
velocity in the $ab$-plane exist. The driving force for
superconductivity is the very strong EPI of these holes with the
$E_{2g}$-phonon mode with the energy of about 70 meV. This mode is
seen as the strong peak in $\Re\Delta _{\sigma }$ in the lower
panel of Fig. 4 \cite{MazAnt}. Note, that the ordinate scale for
the latter
panel is about three times higher than that for the $\pi $-band. For the $\pi -$%
band all the phonons in the energy range $\sim 30\pm 100$ meV take
part
in the $\Re\Delta _{\pi }$ dependence, but with a lower magnitude as compared with the $%
\sigma $-band. Up to about 80 meV $\Im\Delta \left( \omega \right)
$ is small, hence, only $\Re\Delta \left( \omega \right) $ is
important in Eqs. (\ref{bal3}), (\ref{dif3}), (\ref{tun3}). Thus,
we expect that up to $\sim 80$ meV the point-contact spectra
differ only in amplitude just like the
prefactors in the $\left[ \Re\Delta\left( \varepsilon \right) /\varepsilon %
\right] ^{2}$ term. This is evident from Fig. 5.

\begin{figure}[t]
\includegraphics[width=8cm,angle=0]{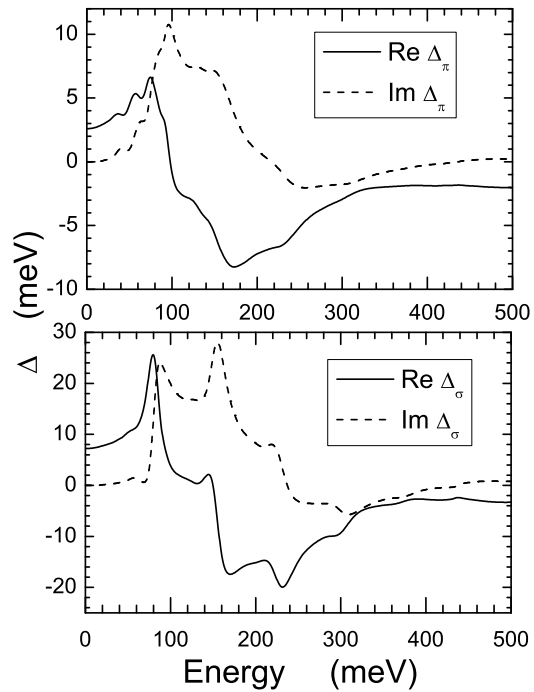}
\caption[]{Real and imaginary parts of $\Delta \left( \varepsilon \right) $%
-dependences for $\pi $- and $\sigma $-bands (upper and lower
panels, respectively) of MgB$_{2}$ \cite{Yan2004}. } \label{fig4}
\end{figure}

We can see that up to the energies $\sim 80\div 90$ meV the shapes
of the calculated spectra are similar (Fig. 5). The discrepancy is
observed only at higher energies, where $\Im\Delta \left( \omega
\right) $ becomes essential. In the range of biases 80$\div $120
meV the variations of point-contact spectra are quite appreciable
and mirror in the experimental curves. Although the $\Delta \left(
\omega \right) $ value is of the order of $\Delta _{\max }$ at
biases higher than 120 meV, the phonon features on the PC spectra
are small enough, due to the factor $1/\varepsilon ^{2}\left(
\omega \right) $. When we compare theoretical and experimental
curves, we consider only the biases up to 120 meV, since for
higher biases the experimental results are irreproducible, due to
nonequilibrium suppression of the superconducting state in the
contact.

\begin{figure}[t]
\includegraphics[width=8cm,angle=0]{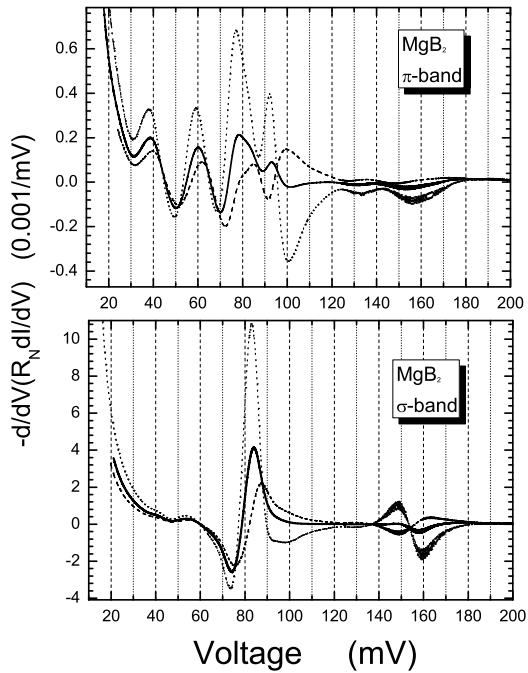}
\caption[]{Calculated point-contact spectra for $\pi $- and
$\sigma $-bands (upper and lower panels, respectively) of
MgB$_{2}$ \cite{Yan2004} for ballistic (dashed curve), diffusive
(thick solid curve) and tunnel (dotted curve) regimes of current
flow. Real and imaginary parts of $\Delta \left( \varepsilon
\right) $ are taken from Fig. 4.} \label{fig5}
\end{figure}

Comparison of calculated and experimental curves is shown in Fig.
6. For the ordinate scale we use the second harmonic voltage
$V_{2}$ measured in experiment, which is related to the second
derivative of the $I-V$ characteristic as follows:

\begin{equation}
V_{2}(V)=-\frac{V_{1}^{2}}{2\sqrt{2}}\frac{d}{dV}\left( R_{N}\frac{dI}{dV}%
\right) ,  \label{V2}
\end{equation}
where $V_{1}$ is the modulation voltage, which is conventionally
taken equal to 3 meV (close to the value used in the experiment).

In the upper panel of Fig. 6 three different point-contact spectra
are shown. They have similar shape up to 100 meV with a slight
increase of
average amplitudes in the row $R_{N}=$ 49$\rightarrow $36$\rightarrow $80 $%
\Omega $. In the same sequence the so-called zero-bias maximum in the $%
dV/dI(V)$ characteristic increases (not shown) , being $R_{0}/R_{N}\leq 1$, $%
\simeq 1$ and $\geq 1$, respectively ($R_{0}$ is the zero-bias
resistance in the superconducting state), for the above mentioned
row. One may speculate that in the same direction the contact
changes from ballistic
towards the tunnel regime\footnote{%
One should be cautioned to ascribe the case $R_{0}/R_{N}\simeq 1$
to the
ballistic regime with the tunnel barrier having $Z$ parameter of about $0.5$%
. In most cases this value of $Z$ in the BTK \cite{BTK} fitting
corresponds to the diffusive regime of current flow. This happens
in many cases, where the same $Z$ value have many junctions with
quite different materials as dissimilar electrodes.}. The lower
panel of Fig. 6 shows that in the same row the average amplitude
of phonon structure increases in the calculated curves.

\begin{figure}[t]
\includegraphics[width=8cm,angle=0]{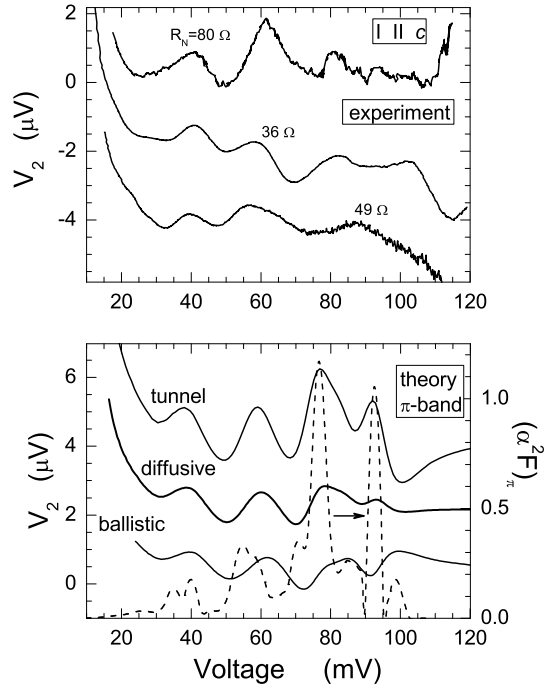}
\caption[]{Comparison of experimental (upper panel) and calculated
(lower panel) phonon structures in point-contact spectra of
MgB$_{2}$. The contact axis is oriented along the $c$-axis. The
normal state resistances are given for each curve. In the
experimental panel the upper curve corresponds to the ordinate
scale, and the other two are shifted down for clarity. $T=4.2$ K.
In the theoretical (lower) panel three transport regimes are
illustrated on the same scale as in the experimental graph. Here,
the lower curve corresponds to the ordinate scale, the other two
are shifted up for clarity. The $\pi$-band EPI function (dashed
curve) \cite{Gol} is shifted to higher voltages by $\Delta
_{\pi}=2.4$ meV. The modulation voltage is taken equal to 3 mV. }
\label{fig6}
\end{figure}

The positions of the maxima of the experimental phonon features
roughly correspond to the expected ones taking into account that
the orientation of experimental junction is approximately along
the $c$-axis, since the spreading behavior of current in direct
conductivity point contact embraces a wide solid
angle near the contact axis. The alignment of our contacts along the $c$%
-axis is inferred indirectly by measuring the Andreev reflection
spectra in the energy gap region and by observation of only the
small gap $\Delta \approx 2.4$ meV characteristic of the $\pi
$-band.

Special attention should be paid to the amplitude of the
experimental curves, which roughly equals that predicted
theoretically. In inelastic spectroscopy the amplitude of the EPI
spectrum is an order of magnitude lower than expected one (see
Fig. 4 in Ref.~\onlinecite{Bob} and the discussion cited therein).
This discrepancy may be explained either by the diffusive regime
of current flow with and unknown mean free path $l_{i}$, or by the
specific PC-transport character of the EPI function obtained from
the inelastic backscattering spectrum.

Comparing theoretical and experimental spectra, one can infer that
all the phonons are essential in the EPI function for the $\pi
$-band. In the lower panel of Fig. 6 we plot the $\left( \alpha
^{2}F\right) _{\pi }$ function taken from Ref. \onlinecite{Gol}.
As expected, the maxima in the EPI spectral function shifted by
$\Delta_{\pi}(0) $ to the right coincide approximately with the
maxima in the second harmonic dependences $V_{2}(V)$.

\section{Conclusion}

The self-energy effect in the phonon feature of a superconducting
point contact can be used, in principle, in the same way as the
Rowell-McMillan program for determination of the EPI spectral
function in tunneling spectroscopy of superconductors
\cite{RowMcM}. Two difficulties arise on this way. One is
theoretical, since this program works well only for the one-band
superconductor, and its application to the two-band case, like
MgB$_{2}$, encounters difficulties \cite{Shulga}. The other is
experimental, since all other sources of $I-V$ nonlinearities
should be removed, and especially, the nonequilibrium effects in
superconductor should be excluded.

The result obtained for superconductivity in MgB$_{2}$ is the
first direct proof  of validity of the calculated EPI function
\cite{Gol}, at least in the $\pi $-band. Unfortunately, this
method encounters difficulty, while applying it to the $\sigma
$-band. In this case the very strong generation of nonequilibrium
phonons cannot be efficiently excluded from the contact region and
this destroys the superconductivity in the constriction.

The question may arise whether the self-energy effects are
important in the normal state. These are known to be smaller than
the inelastic backscattering nonlinearities in the ballistic
regime \cite{Jan}. If we decrease the contact size $d$ or the
elastic mean free path $l_{i}$ in order to make the inelastic
contribution negligible, the latter parameters become comparable
to the Fermi wave length of charge carriers and the strong
nonlinearities connected with localization occur, which masks the
desired phonon structure \cite{Yanloc}.

\section{Acknowledgements}
The author is grateful to Yu.G.~Naidyuk, S.I.~Beloborod'ko,
O.V.~Dolgov and A.A.~Golubov for collaboration. The paper is a
review talk at Miami NATO Advanced Research Workshop on New
Challenges in Superconductivity in January, 2004. The work was
carried out in part by the State Foundation of Fundamental
Research under Grant $\Phi$7/528-2001.


\begin{thebibliography}{}

\bibitem{KOS}  I.O. Kulik, A.N. Omel'yanchuk and R.I. Shekhter, Sov. J. Low
Temp. Phys. \textbf{3}, 840 (1977).

\bibitem{Yan2000}  I.K. Yanson, in book ''\textit{Quantum Mesoscopic
Phenomena and Mesoscopic Devices in Microelectronics'',} I.O.
Kulik and R. Ellialtio\u{g}lu (eds.), p. 61, (2000 ) Kluwer
Academic Publishers.

\bibitem{KulYan}  I.O. Kulik and I.K. Yanson, Sov. J. Low Temp. Phys.
\textbf{4}, 596 (1978).

\bibitem{KSS}  I.O. Kulik, R.I. Shekhter and A.G. Shkorbatov, Sov. Phys.
JETP \textbf{54}, 1130 (1981).

\bibitem{Sha}  Yu. V. Sharvin, Sov. Phys. JETP \textbf{21}, 655 (1965).

\bibitem{OmeKhl}  V.A. Khlus and A.N. Omel'yanchuk , Sov. J. Low Temp. Phys.
\textbf{9}, 189 (1983).

\bibitem{Khl}  V.A. Khlus, Sov. J. Low Temp. Phys. \textbf{9}, 510 (1983).

\bibitem{ArtVolZai}  S.N. Artemenko, A.F. Volkov and A.V. Zaitsev, Solid
State Commun. \textbf{30}, 771 (1979).

\bibitem{BTK}  G.E. Blonder, M. Tinkham and T.M. Klapwijk, Phys. Rev. B
\textbf{25}, 4515 (1982).

\bibitem{OmeBelKul}  A.N. Omel'yanchuk, S.I. Beloborod'ko and V.A. Khlus,
Sov. J. Low Temp. Phys. \textbf{14}, 630 (1988).

\bibitem{BelOme}  S.I. Beloborod'ko and A.N. Omel'yanchuk, Sov. J.
Low Temp. Phys. \textbf{14}, No.3, (1988).

\bibitem{Yan2004}  I.K. Yanson, S.I. Beloborod'ko. Yu.G. Naidyuk, O.V.
Dolgov and A.A. Golubov, Phys. Rev. B (Rapid Communication), to be
published (2004).

\bibitem{RowDyn}  J.M. Rowell, W.L. McMillan and R.C. Dynes, Bell
Laboratories Preprint II, 145 pages, (1969).

\bibitem{Khot}  A.V. Khotkevich, V.V. Khotkevich, I.K. Yanson and G.B.
Kamarchuk, Sov. J. Low Temp. Phys. \textbf{16}, 693 (1990).

\bibitem{MazAnt}  I.I. Mazin and V.P. Antropov, Physica C \textbf{385}, 49
(2003).

\bibitem{Bob}  N.L. Bobrov, P.N. Chubov, Yu.G. Naidyuk, L.V. Tyutrina, I.K.
Yanson, W.N. Kang, H.-J. Kim, E.-M. Choi, and S.-I. Lee, in book \textit{%
''New Trends in Superconductivity'', }J.F. Annett and S. Kruchinin
(eds.), p. 225 (2002) Kluwer Academic Publishers.

\bibitem{Gol}  A.A. Golubov, J. Kortus, O.V. Dolgov, O. Jepsen, Y. Kong,
O.K. Andersen, B. Gibson, K. Ahn, and R.K. Kremer, J. Phys.:
Condens. Matter \textbf{14} 1353 (2002).

\bibitem{RowMcM}  E.L. Wolf, ''\textit{Principles of electron tunneling
spectroscopy'', }Oxford University Press, Inc. New York.

\bibitem{Shulga}  O.V. Dolgov, R.S. Gonnelli, G.A. Ummarino, A.A. Golubov,
S.V. Shulga, and J. Kortus, Phys. Rev. B \textbf{68}, 132503
(2003).

\bibitem{Jan}  A.G.M. Jansen, A.P. van Gelder and P. Wyder, J. Phys. C
\textbf{13}, 6073 (1980).

\bibitem{Yanloc}  I.K. Yanson, O.I. Shklyarevskii and N.N. Gribov, J. Low
Temp. Phys. \textbf{88}, 135 (1992).

\end{thebibliography}
\end{document}